\documentclass[cameraready]{Interspeech}
\usepackage{amsmath}
\usepackage{mathtools}
\usepackage[table]{xcolor}
\usepackage{multirow}
\usepackage{amssymb}
\usepackage{algorithm}
\usepackage{algpseudocode}
\usepackage{arydshln}
\usepackage{makecell}
\usepackage{subcaption}
\usepackage{xcolor}
\definecolor{myblue}{rgb}{0.933, 0.933, 0.996}

\title{Disentangling Speaker Traits for Deepfake Source Verification via \\Chebyshev Polynomial and Riemannian Metric Learning}

\vspace{-1.3cm}
\author[affiliation={1,2}, orcid=0009-0009-5952-1015]{Xi}{Xuan}
\author[affiliation={3}, orcid=0009-0000-8916-6944]{Wenxin}{Zhang}
\author[affiliation={4}, orcid=0009-0004-2020-8148]{Zhiyu}{Li}
\renewcommand{\authorsep}{,\\}
\author[affiliation={5}, orcid=0000-0003-1410-0427]{Jennifer}{Williams}
\renewcommand{\authorsep}{,}
\author[affiliation={1}, orcid=0000-0002-5885-0003]{Ville}{Hautam\"aki}
\renewcommand{\authorsep}{, }
\author[affiliation={1}, orcid=0000-0002-4371-7322]{Tomi H.}{Kinnunen}

\vspace{-2.95cm}
{\footnotesize
\address{
    $^1$ University of Eastern Finland, Finland
    $^2$ University of Illinois Urbana-Champaign, USA\\
    $^5$ University of Southampton, UK
    $^3$ University of Chinese Academy of Sciences, China\\
    $^4$ University of Science and Technology of China, China
    }\email{xi.xuan@uef.fi, J.Williams@soton.ac.uk, ville.hautamaki@uef.fi, tomi.kinnunen@uef.fi}
}


\usepackage{comment}
\definecolor{myheader}{RGB}{70,80,160}

\makeatletter
\def\@maketitle{
    \newpage
    \newbox\titlebox
    \setbox\titlebox=\vbox{%
    \hsize=\textwidth
    \vspace*{-0.5cm}
    \begin{center}
        {\large \bf \@title \par}
        \vskip 6pt
        {
            \large
            \textit{\anonname}
            \vskip 8pt
            \anonaddress
        }
        \vskip 3pt
        {\small\texttt{\anonemail}}
        \par
    \end{center}
    \par
    \thispagestyle{empty}
    \vskip 9pt
    }

    \newdimen\titleboxheight
    \titleboxheight=\ht\titlebox
    \ifdim\titleboxheight<\defaulttitleboxheight
      \titleboxheight=\defaulttitleboxheight
    \fi

    \vbox to \titleboxheight{%
        \unvbox\titlebox
        \vfill
    }
}
\makeatother

\begin{document}
\vspace{-1.9cm}
\maketitle
\vspace{-1.9cm}
\begin{abstract}

Speech deepfake source verification systems aims to determine whether two synthetic speech utterances originate from the same source generator, often assuming that the resulting source embeddings are independent of speaker traits. However, this assumption remains unverified. In this paper, we first investigate the impact of speaker factors on source verification. We propose a speaker-disentangled metric learning (SDML) framework incorporating two novel loss functions. The first leverages Chebyshev polynomial to mitigate gradient instability during disentanglement optimization. The second projects source and speaker embeddings into hyperbolic space, leveraging Riemannian metric distances to reduce speaker information and learn more discriminative source features. Experimental results on MLAAD benchmark, evaluated under four newly proposed protocols designed for source-speaker disentanglement scenarios, demonstrate the effectiveness of SDML framework. The code, evaluation protocols and demo website are available at \footnote{https://github.com/xxuan-acoustics/RiemannSD-Net}.

\end{abstract}

\vspace{-0.20cm}
\section{Introduction}

Speech deepfake generation has made accelerated progress in the last few years~\cite{wani2024navigating,survey,yan2024df40,xuan2024conformer,ZHANG2026132741}, drastically reducing the listeners' perception gap between synthetic and bonafide speech to a point where the difference is barely noticeable~\cite{mai2023warning}. Such high fidelity raises concerns about potential misuse and underscores the urgent need to develop effective countermeasures. Therefore, benchmarks such as the ASVspoof challenges~\cite{wu2017asvspoof,liu2023asvspoof,wang2026asvspoof} promote development of methods for speech deepfake detection. While the deepfake detection task remains important, identifying the source generator (\emph{source tracing}) is equally important. Recent studies \cite{negroni25_interspeech, negroni2025} introduced a scalable, open-set source tracing formulation termed \emph{speech deepfake source verification}, similar to that of speaker verification task. This shifts the focus from identifying the source generator to determining whether a pair of utterances originates from the same or different sources.

Synthetic speech waveforms encode both synthesis traces and high-level factors such as speaking style, prosody, recording conditions, and speaker identity, which are often entangled with the synthesis traces in complex ways~\cite{baade24_interspeech, kassiotis2025disentangling}. However, how these factors, particularly those related to the \emph{speaker}, influence source verification remains largely unexplored. This entanglement allows speaker traits to dominate the embedding space and induce \emph{shortcut learning}~\cite{geirhos2020shortcut}, causing models to rely on speaker cues rather than source evidence. \textbf{Disentangling speaker traits from synthesis source embedding therefore remains an open question and calls for deeper investigation.}

The aim of the present study is to design a well-generalized speaker-disentangled framework for deepfake source verification. At an early stage of our study, we conducted a pilot experiment to probe the relationship between speaker verification and deepfake source verification through a \emph{cross-task evaluation}. Ideally, a model designed for deepfake source verification should not succeed at speaker verification, and a model designed for speaker verification should not succeed at source verification. The results in Table~\ref{tab:cross_task1} demonstrate that this is not the case, indicating that the embeddings learned for each task still retain substantial information about the other task, revealing non-negligible entanglement between speaker and source traits. This motivated us to address more elaborate ways to optimize the embedding  model so that deepfake source verification does not overly rely on speaker cues and remains robust when speaker identity is non-discriminative.

\begin{table}[t]
\centering
\footnotesize
\setlength{\tabcolsep}{3.5pt}

\caption{Pilot experiment on the MLAAD dataset. High cross-task performance between Task1 (speaker verification) and Task2 (source verification) reveals strong entanglement of speaker traits and source, motivating speaker disentanglement.}
\label{tab:cross_task1}
\begin{tabular}{@{}l cc cc@{}}
\toprule[1.5pt]
\multirow{2}{*}{\makecell{\textbf{Embedding}\\\textbf{Extractor}}} &
\multicolumn{2}{c}{\textbf{Task 1}} &
\multicolumn{2}{c}{\textbf{Task 2}} \\
\cmidrule(lr){2-3}\cmidrule(lr){4-5}
& \textbf{EER (\%)} & \textbf{AUC} & \textbf{EER (\%)} & \textbf{AUC} \\
\midrule
Speaker & \cellcolor{myblue}1.85 & \cellcolor{myblue}0.9954 & 29.42 & 0.7845 \\
\hdashline
Source  & 15.34 & 0.9123 & \cellcolor{myblue}5.18 & \cellcolor{myblue}0.9737 \\
\bottomrule[1.5pt]
\end{tabular}
\vspace{-0.4cm}
\end{table}

As the first study on speaker-disentangled deepfake source verification, we propose two novel approaches for disentangling speaker traits and source traces. Both are grounded in the classical theories of renowned mathematicians Pafnuty Chebyshev and Bernhard Riemann: the ChebyAAM loss~\cite{wang2026achilles}, which draws on Chebyshev polynomial approximation~\cite{clenshaw1955note} to enhance training stability, and the HAM-Softmax loss~\cite{fang2026}, which is founded on Riemannian geometry~\cite{chavel1995riemannian} to better model complex distributions of speaker characteristics. Unlike Euclidean space, hyperbolic space can effectively capture the tree-like hierarchical structures of speaker features and synthesis sources~\cite{yang25l_interspeech,shen2010speaker}. Hence, we hypothesize that leveraging the above approaches can better disentangle speaker traits and improve the robustness of speech deepfake source verification systems.

Our study first attempts to design a speaker-disentangling framework by integrating metric learning with polynomial approximation and geometric theory, respectively. We investigate the practical effectiveness of this combined approach, whose impact is previously unexplored in the source verification task. Providing initial answers to this question forms the main novelty of our work. We implement the proposed method using four models, evaluated under four newly proposed protocols designed for diverse source-speaker disentanglement scenarios.

\section{Related Work} 

\subsection{Metric Learning for Deepfake Source Verification} Recent work~\cite{falez25_interspeech} in deepfake source verification has primarily adopted deep metric learning methods. They aim for a compact intra-class distribution and a separated inter-class distribution by adding a margin directly in the angular space to learn representations that discriminate different source generators. For instance, the multi-class N-pair loss~\cite{sohn2016improved} has been integrated with a Conformer~\cite{gulati20_interspeech} and Regmixup~\cite{NEURIPS2022_5ddcfaad} to improve the disentanglement of synthetic sources and overall discriminative ability~\cite{kulkarni25_interspeech}. Furthermore, various metric learning loss functions, including AM-Softmax~\cite{8331118}, AAM-Softmax~\cite{deng2019arcface}, GE2E~\cite{wan2018generalized}, and Angular Prototypical Loss~\cite{chung20b_interspeech}, have been comparatively evaluated for source tracing, demonstrating that AAM-Softmax achieves the best performance~\cite{koutsianos25_interspeech}.

\subsection{Speaker Factor in Speech Deepfakes} 
For the deepfake detection task, recent work~\cite{dao2026assessing} investigated the speaker identity factor and proposed a speaker-invariant multi-task framework incorporating a gradient reversal layer, revealing that the removal of speaker information results in a substantial performance degradation. For the deepfake source verification task, recent work~\cite{negroni25_interspeech} preliminarily explored the impact of speaker distribution. Their findings show that multi-speaker training biases models toward speaker- rather than source-related cues, while single-speaker training encourages focus on source cues. Unlike \cite{dao2026assessing}, which explores the speaker factor in detection tasks, and \cite{negroni25_interspeech}, which investigates the impact of speaker distribution, we focus on designing a speaker-disentangling framework by integrating metric learning with two novel loss functions based on Chebyshev polynomial approximation and Riemannian geometric theory, respectively.

\section{Methodology}

\begin{figure}[t]
    \centering
    \includegraphics[width=0.95\columnwidth]{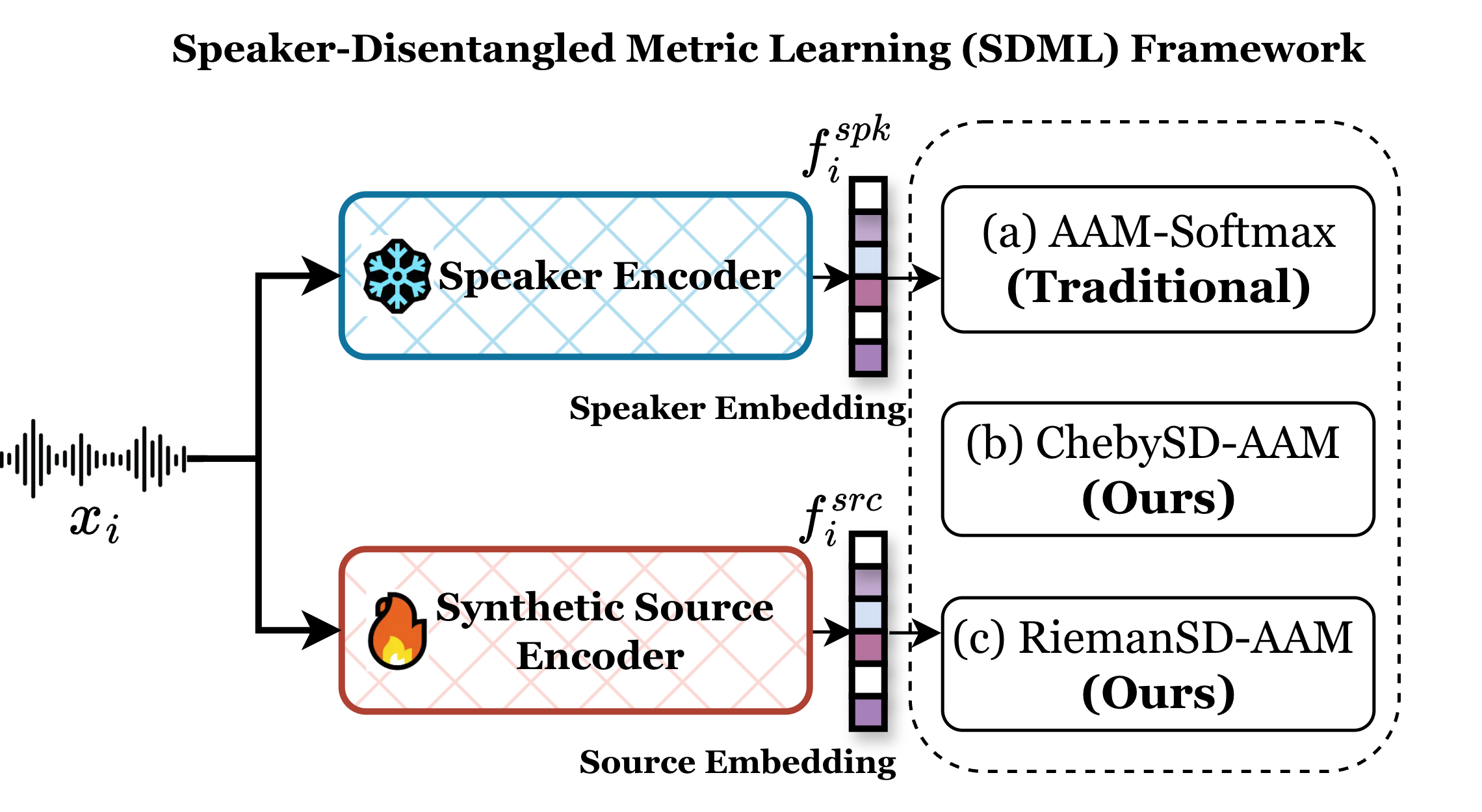}
    
    \caption{Overview of the proposed SDML framework.}
    \label{fig:framework}
    \vspace{-0.4cm}
\end{figure}

This section details the proposed speaker-disentangled metric learning (SDML) framework for source verification task. As shown in Fig.~\ref{fig:framework}, SDML adopts a dual-branch architecture designed to decouple speaker traits from synthesis source representations. For a given input speech utterance $x_i$, a trainable source encoder extracts the deepfake source embedding $f_i^\text{src}$, and a frozen speaker verification model (ReDimNet-B6~\cite{yakovlev2024reshape}) extracts the corresponding speaker embedding $f_i^\text{spk}$. To suppress speaker-related factors during the optimization of $f_i^\text{src}$, we propose two novel loss functions. Before detailing these formulations, we first establish the preliminaries by reviewing the standard AAM-Softmax~\cite{deng2019arcface} and its recent variant, ChebyAAM~\cite{wang2026achilles}.

\begin{table*}[t]
\centering
\caption{Performance comparison of the AAM-Softmax (baseline) with the proposed ChebySD-AAM and RiemannSD-AAM on four proposed evaluation protocols. The best results are in \textbf{bold}, and the second-best are \underline{underlined}. Confidence intervals are in parentheses.}
\vspace{-0.2cm}
\label{tab:ablation_encoder_loss}
\resizebox{\linewidth}{!}{%
\begin{tabular}{l l cc cc cc cc cc}
\toprule[1.5pt]
\multirow{3}{*}{\large\textbf{Encoder}} & \multirow{3}{*}{\makecell{\large\textbf{Loss} \\ \large\textbf{Function}}} & \multicolumn{4}{c}{\rule{0pt}{12pt}\large\textbf{Seen Source}} & \multicolumn{4}{c}{\large\textbf{Unseen Source}} & \multicolumn{2}{c}{\large\textbf{Average}} \\
\cmidrule(lr){3-6} \cmidrule(lr){7-10} \cmidrule(lr){11-12}
& & \multicolumn{2}{c}{\large\textbf{Same Spk (P-I)}} & \multicolumn{2}{c}{\large\textbf{Diff Spk (P-II)}} & \multicolumn{2}{c}{\large\textbf{Same Spk (P-III)}} & \multicolumn{2}{c}{\large\textbf{Diff Spk (P-IV)}} & \multicolumn{2}{c}{\large\textbf{(All Sets)}} \\
\cmidrule(lr){3-4} \cmidrule(lr){5-6} \cmidrule(lr){7-8} \cmidrule(lr){9-10} \cmidrule(lr){11-12}
& & \large\textbf{EER(\%)} $\downarrow$ & \large\textbf{AUC} $\uparrow$ & \large\textbf{EER(\%)} $\downarrow$ & \large\textbf{AUC} $\uparrow$ & \large\textbf{EER(\%)} $\downarrow$ & \large\textbf{AUC} $\uparrow$ & \large\textbf{EER(\%)} $\downarrow$ & \large\textbf{AUC} $\uparrow$ & \large\textbf{EER(\%)} $\downarrow$ & \large\textbf{AUC} $\uparrow$ \\

\toprule[1.5pt]

\multirow{3}{*}{ECAPA-TDNN}
& Baseline & \makecell{0.94 ($\pm$0.05)} & \makecell{0.995 ($\pm$0.002)} & \makecell{1.66 ($\pm$0.08)} & \makecell{0.994 ($\pm$0.003)} & \makecell{6.60 ($\pm$0.21)} & \makecell{0.965 ($\pm$0.008)} & \makecell{11.56 ($\pm$0.35)} & \makecell{0.941 ($\pm$0.012)} & \makecell{5.19 ($\pm$0.17)} & \makecell{0.974 ($\pm$0.006)} \\
& ChebySD-AAM & \makecell{\underline{0.74} ($\pm$0.04)} & \makecell{\underline{0.997} ($\pm$0.001)} & \makecell{\underline{1.22} ($\pm$0.06)} & \makecell{\underline{0.994} ($\pm$0.002)} & \makecell{\underline{5.72} ($\pm$0.18)} & \makecell{\underline{0.972} ($\pm$0.006)} & \makecell{\underline{9.12} ($\pm$0.28)} & \makecell{\underline{0.946} ($\pm$0.010)} & \makecell{\underline{4.20} ($\pm$0.14)} & \makecell{\underline{0.977} ($\pm$0.005)} \\
\rowcolor{myblue}
& RiemannSD-AAM & \makecell{\textbf{0.73} ($\pm$0.03)} & \makecell{\textbf{0.998} ($\pm$0.001)} & \makecell{\textbf{1.20} ($\pm$0.05)} & \makecell{\textbf{0.995} ($\pm$0.002)} & \makecell{\textbf{5.53} ($\pm$0.16)} & \makecell{\textbf{0.980} ($\pm$0.005)} & \makecell{\textbf{9.06} ($\pm$0.27)} & \makecell{\textbf{0.952} ($\pm$0.008)} & \makecell{\textbf{4.13} ($\pm$0.13)} & \makecell{\textbf{0.981} ($\pm$0.004)} \\
\hdashline

\multirow{3}{*}{ResNet34}
& Baseline & \makecell{0.73 ($\pm$0.04)} & \makecell{0.997 ($\pm$0.002)} & \makecell{1.38 ($\pm$0.07)} & \makecell{0.994 ($\pm$0.003)} & \makecell{7.24 ($\pm$0.22)} & \makecell{0.971 ($\pm$0.007)} & \makecell{9.77 ($\pm$0.29)} & \makecell{0.962 ($\pm$0.009)} & \makecell{4.78 ($\pm$0.15)} & \makecell{0.981 ($\pm$0.005)} \\
& ChebySD-AAM & \makecell{\underline{0.71} ($\pm$0.04)} & \makecell{\underline{0.998} ($\pm$0.001)} & \makecell{1.35 ($\pm$0.06)} & \makecell{\underline{0.995} ($\pm$0.002)} & \makecell{\underline{5.85} ($\pm$0.19)} & \makecell{\underline{0.974} ($\pm$0.006)} & \makecell{\underline{8.24} ($\pm$0.25)} & \makecell{\underline{0.969} ($\pm$0.007)} & \makecell{\underline{4.04} ($\pm$0.13)} & \makecell{\underline{0.984} ($\pm$0.004)} \\
\rowcolor{myblue}
& RiemannSD-AAM & \makecell{\textbf{0.68} ($\pm$0.03)} & \makecell{\textbf{0.998} ($\pm$0.001)} & \makecell{\textbf{1.21} ($\pm$0.05)} & \makecell{\textbf{0.996} ($\pm$0.001)} &  \makecell{\textbf{4.08} ($\pm$0.14)}  & \makecell{\textbf{0.988} ($\pm$0.003)} & \makecell{\textbf{7.13} ($\pm$0.21)} & \makecell{\textbf{0.972} ($\pm$0.006)} & \makecell{\textbf{3.27} ($\pm$0.10)} & \makecell{\textbf{0.988} ($\pm$0.003)} \\
\hdashline

\multirow{3}{*}{AASIST}
& Baseline & \makecell{1.20 ($\pm$0.06)} & \makecell{0.992 ($\pm$0.003)} & \makecell{1.58 ($\pm$0.08)} & \makecell{0.990 ($\pm$0.004)} & \makecell{7.46 ($\pm$0.24)} & \makecell{0.972 ($\pm$0.007)} & \makecell{12.26 ($\pm$0.38)} & \makecell{0.935 ($\pm$0.013)} & \makecell{5.62 ($\pm$0.19)} & \makecell{0.972 ($\pm$0.006)} \\
& ChebySD-AAM & \makecell{\underline{0.89} ($\pm$0.05)} & \makecell{\underline{0.993} ($\pm$0.002)} & \makecell{\underline{1.47} ($\pm$0.07)} & \makecell{\underline{0.990} ($\pm$0.004)}  & \makecell{\underline{5.25} ($\pm$0.16)} & \makecell{\underline{0.982} ($\pm$0.005)} & \makecell{\underline{10.93} ($\pm$0.33)} & \makecell{\underline{0.953} ($\pm$0.009)} & \makecell{\underline{4.64} ($\pm$0.15)} & \makecell{\underline{0.979} ($\pm$0.005)} \\
\rowcolor{myblue}
& RiemannSD-AAM  & \makecell{\textbf{0.79} ($\pm$0.04)} & \makecell{\textbf{0.993} ($\pm$0.002)} & \makecell{\textbf{1.42} ($\pm$0.06)} & \makecell{\textbf{0.991} ($\pm$0.003)}  & \makecell{\textbf{4.41} ($\pm$0.14)} &  \makecell{\textbf{0.985} ($\pm$0.004)}  & \makecell{\textbf{9.82} ($\pm$0.29)} & \makecell{\textbf{0.961} ($\pm$0.007)} & \makecell{\textbf{4.11} ($\pm$0.13)} & \makecell{\textbf{0.982} ($\pm$0.004)} \\
\hdashline

\multirow{3}{*}{Mamba}
& Baseline & \makecell{1.31 ($\pm$0.07)} & \makecell{0.994 ($\pm$0.003)} & \makecell{1.81 ($\pm$0.09)} & \makecell{0.992 ($\pm$0.003)} & \makecell{9.54 ($\pm$0.28)} & \makecell{0.926 ($\pm$0.015)} & \makecell{13.93 ($\pm$0.42)}   & \makecell{0.928 ($\pm$0.014)} & \makecell{6.65 ($\pm$0.21)} & \makecell{0.960 ($\pm$0.010)} \\
& ChebySD-AAM & \makecell{\underline{0.81} ($\pm$0.04)} & \makecell{\underline{0.994} ($\pm$0.002)} & \makecell{\underline{1.47} ($\pm$0.07)} & \makecell{\underline{0.994} ($\pm$0.002)} & \makecell{\underline{6.68} ($\pm$0.21)} & \makecell{\underline{0.969} ($\pm$0.007)} & \makecell{\underline{10.96} ($\pm$0.34)} & \makecell{\underline{0.951} ($\pm$0.010)} & \makecell{\underline{4.98} ($\pm$0.16)} & \makecell{\underline{0.977} ($\pm$0.006)} \\
\rowcolor{myblue}
& RiemannSD-AAM & \makecell{\textbf{0.75} ($\pm$0.03)} & \makecell{\textbf{0.996} ($\pm$0.001)} & \makecell{\textbf{1.29} ($\pm$0.06)} & \makecell{\textbf{0.995} ($\pm$0.002)} & \makecell{\textbf{4.59} ($\pm$0.15)} & \makecell{\textbf{0.982} ($\pm$0.004)} & \makecell{\textbf{9.48} ($\pm$0.28)} & \makecell{\textbf{0.964} ($\pm$0.007)} & \makecell{\textbf{4.02} ($\pm$0.12)} & \makecell{\textbf{0.984} ($\pm$0.004)} \\

\bottomrule[1.5pt]
\vspace{-0.8cm}
\end{tabular}}
\end{table*}

\subsection{Background: AAM-Softmax and Cheby-AAM}

The standard AAM-Softmax~\cite{deng2019arcface} is designed to enhance intra- class compactness and inter-class discrepancy. For the $i$-th sample, 
it is defined by
\begin{equation}
\label{eq:aam_softmax}
L_{\text{AAM}} = -\log \frac{e^{s.\cos(\theta_{y_i} + m)}}{e^{s.\cos(\theta_{y_i} + m)} + \sum_{j \neq y_i} e^{s.\cos(\theta_j)}},
\end{equation}
where $\theta_{y_i}$ denotes the angle between a source embedding $f_i^\text{src}$ and the target class weight vector (prototype) $W_{y_i}^\text{src}$, $s$ denotes the scale factor, and $m$ is an additive margin control parameter.

Despite remaining popular in metric learning applications, the standard AAM-Softmax~\eqref{eq:aam_softmax} suffers from two major shortcomings, as identified in ~\cite{wang2026achilles}:
\emph{gradient instability} and \emph{insufficient penalisation of hard examples}. To elaborate, note that optimizing \eqref{eq:aam_softmax} requires evaluation of the derivative of the arccos function, which becomes unbounded at $\pm 1$. During optimization, this leads to gradient instability once the learnt embeddings approach their class prototypes. This observation motivated the authors in~\cite{wang2026achilles} to replace the composite function $\cos(\arccos(x) + m)$ with its \emph{Chebyshev polynomial approximation}, i.e. 
$\mathcal{F}_{\text{cheb}}(x, m) = \frac{1}{2}a_0 + \sum_{k=1}^{K} a_k T_k(x)$, where the 
coefficients $a_k$ are given in ~\cite[Eq.(2)]{wang2026achilles}, and $T_k(x) = \cos(k \arccos(x))$. Chebyshev polynomials are well suited for approximating functions on the interval $[-1,1]$, as they provide stable approximations with uniformly controlled error across the entire interval. This helps 
mitigating both the gradient explosion problem and providing a stronger corrective signal for hard examples.

\subsection{Proposed ChebySD-AAM}

Whereas both AAM-Softmax and ChebyAAM encourage angular separation of different deepfake source generators, neither is designed to cope with entangled speaker-related effects. Our first speaker-disentanglement loss, \textbf{Chebyshev} \underline{\textbf{S}}peaker \underline{\textbf{D}}isentangled-\underline{\textbf{AAM}} Softmax (\textbf{ChebySD-AAM}), extends ChebyAAM~\cite{wang2026achilles} by including additional speaker margin terms. Specifically, we introduce a \textit{thresholded speaker adaptive margin}, denoted as $\mathcal{M}_{i}^{\text{spk}}$, to adjust the decision boundary. 
The new loss is formulated as:
\begin{equation}
\label{eq:cheby_sd_aam}
L= -\log \frac{e^{s \cdot \mathcal{F}_{\text{cheb}}(\cos\theta_{y_i},\, m)}}{e^{s \cdot \mathcal{F}_{\text{cheb}}(\cos\theta_{y_i},\, m)} + \sum_{j \neq y_i} e^{s\cdot(\cos\theta_j + \lambda \mathcal{M}_{i}^{\text{spk}})}},
\end{equation}

\noindent where $\cos\theta_{y_i} = (W_{y_i}^\text{src})^\top f_i^{\text{src}}$ is the cosine similarity between the source embedding and the prototype of generator $j$. The numerator applies the Chebyshev approximation $\mathcal{F}_{\text{cheb}}(\cdot, m)$ of degree $K$ to the target class similarity, preserving the gradient stability of ChebyAAM. A lower $K$ produces a smoother margin with stronger regularization, while a higher $K$ more closely approximates the original angular margin function. Each non-target logit is augmented by $\lambda \mathcal{M}_{i}^{\text{spk}}$, where $\lambda$ is a disentanglement coefficient and $\mathcal{M}_{i}^{\text{spk}} = \max\big(0,\; |(f_i^{\text{src}})^\top f_i^{\text{spk}}| - \tau\big)$ penalizes alignment between source and speaker embeddings. $\mathcal{M}_{i}^{\text{spk}}$ becomes positive and raises the non-target logits, reducing the score gap between target and non-target classes. This steers the encoder toward learning source representations with reduced speaker information. The parameters $K$ and $\lambda$ are set as fixed values, where $K$ determines the Chebyshev polynomial degree and $\lambda$ serves as the speaker disentanglement coefficient. Their impact is addressed in the ablation study (Section~\ref{subsec:ablation_study}).

\subsection{Proposed  RiemannSD-AAM}

ChebySD-AAM relies on Euclidean geometry to model the relationships between synthesis source classes. Our second speaker disentanglement strategy goes beyond this to consider non-Euclidean (curved) embedding space geometries, motivated from two different perspectives. First, some evidence~\cite{xuan2026wst,10.24963/ijcai.2024/46,SONAR,10516609} points to deepfake artifacts being manifested in the subtle correlations of feature channels (second-order statistics). Second, synthesis source distributions can be assumed to possess tree-like hierarchical structure~\cite{yang25l_interspeech, sheth2025curved}. 
These motivate us to adopt Riemannian geometry (hyperbolic space), informally considered as a continuous analogue of discrete trees and hence suitable for learning intrinsic hierarchical structure of synthesis traces. Unlike Euclidean spaces, Riemannian distance is defined as the minimum length among curves connecting two points in a connected Riemannian manifold~\cite{lin2008riemannian}.

Concretely, our \textbf{RiemannSD-AAM} loss function is inspired by~\cite{fang2026} originally proposed for speaker verification. We project the source 
embedding $f_i^{\text{src}}$ and class prototype $W_{y_i}^\text{src}$ onto the so-called
\emph{Poincar\'{e} ball} via the exponential map at the origin, obtaining 
$\tilde{f}_i^{\text{src}} = \text{proj}(f_i^{\text{src}})$ and 
$\tilde{W}_{y_i}^{\text{src}} = \text{proj}(W_{y_i}^\text{src})$. The hyperbolic distance 
between them is denoted as $d_{i,j} = d_{\mathcal{H}}(\tilde{f}_i^{\text{src}},\, 
\tilde{W}_{y_i}^{\text{src}})$. The new loss is formulated as:
\vspace{-0.1cm}
\begin{equation}
\label{eq:rieman_sd_aam}
L_{\text{RiemannSD-AAM}} = -\log \frac{e^{-s(d_{i,y_i} + m)}}
{e^{-s(d_{i,y_i} + m)} + \displaystyle\sum_{j \neq y_i} 
e^{-s \cdot d_{i,j} + \lambda \mathcal{M}_{i}^{H}}},
\end{equation}
\noindent where $\lambda$ is a disentanglement coefficient and where 
$\mathcal{M}_{i}^{H} = \max\big(0,\; \gamma - 
d_{\mathcal{H}}(\tilde{f}_i^{\text{src}},\, \tilde{f}_i^{\text{spk}})\big)$ is the speaker margin in 
the hyperbolic space, 
$\tilde{f}_i^{\text{spk}} = \text{proj}(f_i^{\text{spk}})$ denoting the frozen speaker 
embedding projected onto the same Poincar\'{e} ball. In hyperbolic space, a small 
distance $d_{\mathcal{H}}(\tilde{f}_i^{\text{src}}, \tilde{f}_i^{\text{spk}})$ 
indicates that the source embedding lies close to the speaker embedding, signaling 
identity leakage. When this distance falls below $\gamma$, $\mathcal{M}_{i}^{H}$ 
becomes positive and raises the non-target logits. The hyperparameters $c$ and 
$\lambda$ are set as fixed values, where $c$ determines the curvature of the 
Poincar\'{e} ball and $\lambda$ serves as the speaker disentanglement coefficient. 
Their selection is discussed in the ablation study 
(Section~\ref{subsec:ablation_study}).
\vspace{-0.2cm}
\section{Experimental Setup}
\label{sec:Experiment}

\subsection{Dataset \& Evaluation Metrics}
We adopt the MLAAD v8 dataset~\cite{muller2024mlaad} for the deepfake source verification task. 
The corresponding official MLAAD protocol ~\cite{mullerusing} consists of 11,100 training, 12,000 development, and 33,900 evaluation samples, covering 38 languages and 82 TTS models across 33 different architectures, totaling 378 hours of synthetic speech. For more details, please refer to~\cite{mullerusing}. Following~\cite{negroni25_interspeech}, the selected performance metrics include EER and AUC. To ensure statistical reliability, we report two times the standard deviation for all metrics using 1,000 bootstrap runs \cite{efron1986bootstrap} on the test dataset.

\vspace{-0.2cm}
\subsection{Protocols for Source-Speaker Disentanglement}
\noindent\textbf{Train.} We followed the official protocols~\cite{mullerusing} and used the combined train and dev sets (23,100 samples) for training.

\noindent\textbf{Evaluation Design.} To evaluate how well our framework disentangles speaker traits from source traces, we adopt a trial-based protocol common in speaker verification~\cite{xuan2024conformer,xuanasv1,xuanasv2,xuanasv3,xuanasv4}. Each evaluation set consists of sample pairs categorized by two primary factors: the generator source and the speaker identity.

\begin{itemize}
    \item \textbf{Source Pairs:} Pairs are classified as \textit{Seen-Seen} if the generators were encountered during training, or \textit{Unseen-Unseen} if they originate from entirely new generators.
    \item \textbf{Speaker Pairs:} Since the MLAAD metadata is faced with the limitation that it does not contain speaker labels. Moreover, arguably, synthetic speech does not even \emph{have} a crisply defined speaker identity, but only a targeted identity used at the training or adaptation stage of text-to-speech or voice conversion systems. Following \cite[Section 3.3]{klein2024source} and \cite[Section 2.2.4]{xuan25_spsc}, we leverage \emph{pseudo-speaker} labels via cosine similarity between speaker embeddings. While an initial threshold of 0.3 was considered, further analysis of the score distribution revealed a distinct `valley' at $\sim$0.5, which we selected as the threshold to obtain binary (same/different speaker) trial keys.

\end{itemize}

\noindent\textbf{Proposed Evaluation Protocols.} By combining these source and speaker conditions, we establish four distinct protocols (P-I to P-IV) to test both disentanglement and generalization. The detailed statistics are presented in Table~\ref{tab:protocols}.

\vspace{-0.3cm}
\begin{table}[H] 
\centering
\caption{Statistics for the different evaluation protocols. All protocols are balanced to ensure a 1:1 positive-negative ratio.}
\vspace{-0.2cm}
\label{tab:protocols}
\resizebox{\linewidth}{!}{
\begin{tabular}{lccrrr}
\toprule[1.5pt]
\textbf{Eval.} & \textbf{Source} & \textbf{Speaker} & \textbf{\# Total} & \textbf{\# Positive} & \textbf{\# Negative} \\
\textbf{Protocol}   & \textbf{Pair}   & \textbf{Pair}    & \textbf{Utterances}      & \textbf{Pairs}       & \textbf{Pairs} \\
\midrule[1.5pt]
\textbf{P-I}   & Seen-Seen     & Same      & 27,530 & 13,765 & 13,765 \\
\textbf{P-II}  & Seen-Seen     & Different & 27,530 & 13,765 & 13,765 \\
\hdashline 
\textbf{P-III} & Unseen-Unseen & Same      & 27,530 & 13,765 & 13,765 \\
\textbf{P-IV}  & Unseen-Unseen & Different & 27,530 & 13,765 & 13,765 \\
\bottomrule[1.7pt]
\vspace{-0.8cm}
\end{tabular} 
}
\end{table}

\vspace{-0.3cm}
\subsection{Implementation Details}

We implement our proposed framework using PyTorch Lightning~\cite{falcon2019pytorch} and SpeechBrain~\cite{ravanelli2024open}. Using Librosa~\cite{mcfee2015librosa}, all audio samples are down-sampled to 16 kHz, with a 3s segment extracted from each utterance. The front-end opts for more stable hand-crafted acoustic features in the form of 80-dimensional linear filterbanks, extracted using a 25 ms Hanning window with a 10 ms frame shift. We compare several representative models as source encoders, including ECAPA-TDNN~\cite{desplanques20_interspeech,xuan2024efficient}, ResNet34~\cite{he2016deep}, AASIST~\cite{jung2022aasist}, and Mamba~\cite{xuan2025fakemamba,xuan2025wavesp}. To mitigate distribution shift between the training and testing data, we applied data augmentation to the training data using additive noise sampled from MUSAN~\cite{snyder2015musan} along with room impulse responses (RIRs)~\cite{ko2017study}. Following~\cite{koutsianos25_interspeech}, we set $s=30$ and $m=0.3$ for AAM-Softmax. We set $\tau{=}0.1$ (ChebySD-AAM) and $\gamma{=}2$ (RiemannSD-AAM). The hyper-parameters $K$, $\lambda$, and $c$ were tuned on the dev set via grid search over $K \in \{5, 10, 20\}$, $\lambda \in \{0.1, 1, 10\}$, and $c \in \{0.5, 3, 6, 10\}$; Table ~\ref{table_ablation_resnet34_sd} reports the parameter sensitivity experiment. We use the Adam optimizer with an initial learning rate of $10^{-3}$, which decays by 10\% every epoch. We also set the weight decay to $10^{-7}$ to avoid overfitting and perform a linear warmup for the first 2k steps. The batch size is 200. We use cosine similarity for scoring.

\begin{figure}[t]
    \centering
    \begin{subfigure}[b]{0.45\columnwidth}
        \centering
        \includegraphics[width=\linewidth]{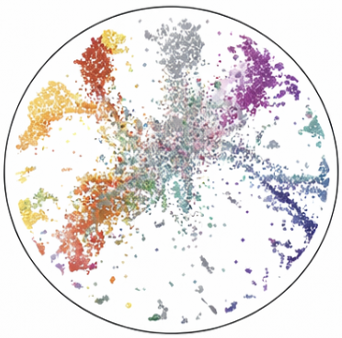}
        \caption{\scriptsize Colored by source ID}
        \label{fig:sne1_c}
    \end{subfigure}\hspace{0.01\columnwidth}
    \begin{subfigure}[b]{0.45\columnwidth}
        \centering
        \includegraphics[width=\linewidth]{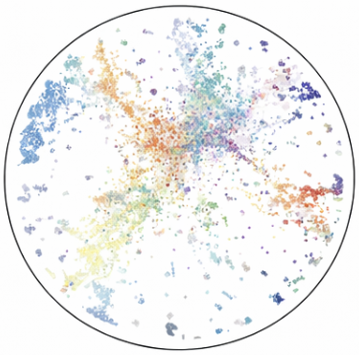}
        \caption{\scriptsize Colored by speaker ID}
        \label{fig:sne1_d}
    \end{subfigure}
    \vspace{-0.3cm}
    \caption{t-SNE visualization of embeddings learned using RiemannSD-AAM.}
    \label{fig:sne1}
    \vspace{-0.7cm}
\end{figure}

\vspace{-0.2cm}
\section{Results}

\vspace{-0.1cm}
\subsection{Framework with Different Loss Functions}
Table \ref{tab:ablation_encoder_loss} shows the performance of our proposed loss functions, ChebySD- and RiemannSD-AAM, compared to the AAM-Softmax across four evaluation protocols. The results indicate that the proposed speaker-disentangled loss functions consistently outperform the baseline (AAM-Softmax), regardless of the synthetic source encoder employed. Notably, the ResNet34 encoder achieves the best overall performance among the evaluated architectures, similar to findings in previous work~\cite{klein24_interspeech,koutsianos25_interspeech}. When paired with this encoder, RiemannSD-AAM yields the best overall results. Its effectiveness demonstrates that by relying on hyperbolic space to model hierarchical structures, while incorporating Euclidean space margin constraints to enhance local discriminability, the model achieves stronger robustness and generalization ability. Correspondingly, ChebySD-AAM also presents competitive results under the same configuration. Its performance reveals that the use of Chebyshev approximation serves to generate a steeper gradient for hard examples, providing a stronger corrective signal where it is most needed and leading to more effective optimization in source verification.

\begin{table}[t]
    \centering
    \caption{\footnotesize Ablation and Parameter Sensitivity with ResNet34 source encoder
    on unseen source protocols. (\checkmark) / ($\times$) denotes whether speaker disentanglement (SD) is ablated. Best in \textbf{bold}, second best \underline{underlined}.
    ChebySD-AAM ($K{=}10$, $\lambda{=}1$); RiemannSD-AAM ($c{=}6$, $\lambda{=}1$).}
    \label{table_ablation_resnet34_sd}
    \vspace{5pt}
    \vspace{-0.4cm}
    \setlength{\tabcolsep}{3pt} 
    \renewcommand{\arraystretch}{1.05}
    \resizebox{\columnwidth}{!}{
    \begin{tabular}{@{}l cccc@{}}
        \Xhline{1.2pt}
& \multicolumn{4}{c}{\rule{0pt}{12pt}\large\textcolor{myheader}{\textbf{Unseen Source}}} \\
\Xhline{0.3pt}
        & \multicolumn{2}{c}{\textbf{Same Spk (P-III)}} & \multicolumn{2}{c}{\textbf{Diff Spk (P-IV)}} \\
        \Xhline{0.3pt}
        \textbf{Method}
        & \textbf{EER(\%)}$\downarrow$ & \textbf{AUC}$\uparrow$
        & \textbf{EER(\%)}$\downarrow$ & \textbf{AUC}$\uparrow$ \\
        \midrule
        Baseline~\cite{deng2019arcface} ($\times$)  & \makecell{7.24 ($\pm$0.22)} & \makecell{0.971 ($\pm$0.007)} & \makecell{9.77 ($\pm$0.29)} & \makecell{0.962 ($\pm$0.009)} \\
        \midrule
        ChebyAAM~\cite{wang2026achilles} ($\times$) & \makecell{6.10 ($\pm$0.20)} & \makecell{0.973 ($\pm$0.006)} & \makecell{8.87 ($\pm$0.27)} & \makecell{0.965 ($\pm$0.008)} \\
        \rowcolor{myblue}
        ChebySD-AAM ($\checkmark$)                  & \makecell{\underline{5.85} ($\pm$0.19)} & \makecell{\underline{0.974} ($\pm$0.006)} & \makecell{\underline{8.24} ($\pm$0.25)} & \makecell{\underline{0.969} ($\pm$0.007)} \\
        \midrule
        HAM-Softmax~\cite{fang2026} ($\times$)      & \makecell{4.53 ($\pm$0.15)} & \makecell{0.985 ($\pm$0.004)} & \makecell{7.48 ($\pm$0.22)} & \makecell{0.970 ($\pm$0.007)} \\
        \rowcolor{myblue}
        RiemannSD-AAM ($\checkmark$)                 & \makecell{\textbf{4.08} ($\pm$0.14)} & \makecell{\textbf{0.988} ($\pm$0.003)} & \makecell{\textbf{7.13} ($\pm$0.21)} & \makecell{\textbf{0.972} ($\pm$0.006)} \\
        \Xhline{1.4pt}
        \multicolumn{5}{c}{\rule{0pt}{12pt}\large\textcolor{myheader}{\textbf{ChebySD-AAM}}} \\
        \Xhline{0.8pt}
        \multicolumn{5}{c}{\textbf{Hyperparameter 1: Polynomial Degree $K$}} \\
        \Xhline{0.3pt}
        $K = 5$  & \makecell{5.96 ($\pm$0.21)} & \makecell{0.972 ($\pm$0.007)} & \makecell{8.37 ($\pm$0.27)} & \makecell{0.967 ($\pm$0.008)} \\
        $K = 20$ & \makecell{5.91 ($\pm$0.20)} & \makecell{0.973 ($\pm$0.007)} & \makecell{8.31 ($\pm$0.26)} & \makecell{0.968 ($\pm$0.007)} \\
        \Xhline{0.8pt}
        \multicolumn{5}{c}{\textbf{Hyperparameter 2: Disentanglement Coefficient $\lambda$}} \\
        \Xhline{0.3pt}
        $\lambda = 0.1$ & \makecell{6.14 ($\pm$0.22)} & \makecell{0.971 ($\pm$0.008)} & \makecell{8.53 ($\pm$0.28)} & \makecell{0.966 ($\pm$0.008)} \\
        $\lambda = 10$  & \makecell{6.07 ($\pm$0.21)} & \makecell{0.972 ($\pm$0.007)} & \makecell{8.44 ($\pm$0.27)} & \makecell{0.967 ($\pm$0.008)} \\
        \Xhline{1.4pt}
        \multicolumn{5}{c}{\rule{0pt}{12pt}\large\textcolor{myheader}{\textbf{RiemannSD-AAM}}} \\
        \Xhline{0.3pt}
        \multicolumn{5}{c}{\textbf{Hyperparameter 1: Curvature $c$}} \\
        \Xhline{0.3pt}
        $c = 0.5$ & \makecell{4.19 ($\pm$0.15)} & \makecell{0.986 ($\pm$0.004)} & \makecell{7.28 ($\pm$0.23)} & \makecell{0.970 ($\pm$0.007)} \\
        $c = 3$   & \makecell{4.11 ($\pm$0.14)} & \makecell{0.987 ($\pm$0.004)} & \makecell{7.19 ($\pm$0.22)} & \makecell{0.971 ($\pm$0.006)} \\
        $c = 10$  & \makecell{4.24 ($\pm$0.16)} & \makecell{0.986 ($\pm$0.004)} & \makecell{7.37 ($\pm$0.24)} & \makecell{0.970 ($\pm$0.007)} \\
        \Xhline{0.8pt}
        \multicolumn{5}{c}{\textbf{Hyperparameter 2: Disentanglement Coefficient $\lambda$}} \\
        \Xhline{0.3pt}
        $\lambda = 0.1$ & \makecell{4.17 ($\pm$0.15)} & \makecell{0.987 ($\pm$0.004)} & \makecell{7.24 ($\pm$0.22)} & \makecell{0.971 ($\pm$0.006)} \\
        $\lambda = 10$  & \makecell{4.29 ($\pm$0.16)} & \makecell{0.985 ($\pm$0.005)} & \makecell{7.41 ($\pm$0.24)} & \makecell{0.969 ($\pm$0.007)} \\
        \Xhline{1.4pt}
    \end{tabular}}
    \vspace{-0.6cm}
\end{table}

\vspace{-0.2cm}
\subsection{Ablation \& Parameter Sensitivity Experiments}
\label{subsec:ablation_study}
Table~\ref{table_ablation_resnet34_sd} provides ablation results within the SDML 
framework, showing that removing speaker disentanglement leads to notable performance 
degradation, confirming its necessity in the challenging unseen source scenarios. For ChebySD-AAM, a moderate polynomial degree 
$K$ provides sufficient approximation capacity without overfitting, while an intermediate 
disentanglement coefficient $\lambda$ strikes the best balance between penalizing speaker 
entanglement and preserving source discriminability. For RiemannSD-AAM, better performance 
is achieved at higher curvatures ($c=6$), confirming that high-curvature hyperbolic space 
better models the hierarchical structure of source distributions; 
similarly, an intermediate $\lambda$ balances the reduction of speaker information and the preservation of source-discriminative structure.

\vspace{-0.2cm}
\subsection{Visualization of Embedding Disentanglement}
Figure~\ref{fig:sne1} visualizes the embeddings learned by RiemannSD-AAM using t-SNE. When colored by source ID (Fig.~\ref{fig:sne1}(a)), the embeddings form distinct and 
compact clusters, indicating effective inter-class separation. Conversely, when colored 
by speaker ID (Fig.~\ref{fig:sne1}(b)), the embeddings are dispersed without 
discernible clusters, demonstrating successful disentanglement of speaker traits from deepfake source representations.

\vspace{-0.2cm}
\section{Conclusion and Future Work}

For the first time, our study addressed speaker disentanglement in speech deepfake source verification. Through a pilot cross-task evaluation experiment, we first revealed an entanglement between speaker traits and the synthesis source in existing source verification systems, demonstrating that source embeddings encode speaker-related information. Motivated by this finding, we introduced two novel loss functions, ChebySD-AAM and RiemannSD-AAM, to learn more robust and speaker-invariant source embeddings. Extensive experiments on the recent MLAAD benchmark, conducted across four encoder architectures and four newly proposed evaluation protocols targeting diverse source-speaker disentanglement scenarios, confirm the SDML framework's generalizability.

Future research will extend the SDML framework to isolate and control not only speaker identity but also other entangled attributes, such as speaking style, prosody, language, accent, and recording conditions, towards achieving more flexible and interpretable deepfake source representations.

\vspace{-0.2cm}

\section{Acknowledgment}
This work was supported by the Finnish AI-DOC project “Explainable Speech Deepfake Characterization” (Decision No. VN/3137/2024-OKM-6), and the Research Council of Finland, project “SPEECHFAKES” (Decision No. 349605).

\section{Generative AI Use Disclosure}
Generative AI was used to check grammatical errors, shortening texts and editing LaTeX more efficiently. All authors reviewed and approved the manuscript before submission.

\bibliographystyle{IEEEtran}
\bibliography{mybib}

\end{document}